\documentclass[twocolumn,superscriptaddress,floatfix,prl,showpacs]{revtex4}
\usepackage{graphicx}

 \newcommand{\mb}[1]{\mathbf{#1}}

\begin{document}
\title{Mobility of Holstein polaron: an unbiased approach}
\author{A.~S.~Mishchenko}
\affiliation{RIKEN Center for Emergent Matter Science (CEMS),  
2-1 Hirosawa, Wako, Saitama, 351-0198, Japan}
\affiliation{RRC ``Kurchatov Institute", 123182, Moscow, Russia}

\author{N.~Nagaosa}
\affiliation{RIKEN Center for Emergent Matter Science (CEMS),
2-1 Hirosawa, Wako, Saitama, 351-0198, Japan}
\affiliation{Department of Applied Physics, The University of Tokyo,
7-3-1 Hongo, Bunkyo-ku, Tokyo 113, Japan} 

\author{G.~De~Filippis}
\affiliation{SPIN-CNR and Dip. di Fisica - 
Universit\`{a} di Napoli Federico II - I-80126 Napoli, Italy}

\author{A.~de~Candia}
\affiliation{SPIN-CNR and Dip. di Fisica - 
Universit\`{a} di Napoli Federico II - I-80126 Napoli, Italy}

\author{V.~Cataudella}
\affiliation{SPIN-CNR and Dip. di Fisica - 
Universit\`{a} di Napoli Federico II - I-80126 Napoli, Italy}


\begin{abstract}

We present the first unbiased results for the mobility $\mu$ of one-dimensional Holstein polaron
obtained by numerical analytic continuation combined with diagrammatic and world-line 
Monte Carlo methods in the thermodynamic limit.
We have identified for the first time, by the characteristic $\omega$ and $T$ dependence
in the wide region of parameters, several distinct regimes in the  $\lambda-T$ plane 
including band conduction region, incoherent metallic region, 
activated hopping region, and high temperature saturation region.
We observe for the first time that although mobilities and mean free paths at different values 
of $\lambda$ differ by many orders of magnitude at small temperatures, their values at 
$T$ larger than the bandwidth become very close to each other.     
\end{abstract}

\maketitle

Motion of a quantum mechanical particle in a background of quantum phonons 
(polaron) is an issue of fundamental importance both theoretically and experimentally
\cite{Alex}. 
Historically, it is the first condensed matter problem where the quantum field 
theory has been applied successfully \cite{FHIP}.
 Although it is a single-particle problem, 
many quantum phonon modes are involved and the many-body nature of the 
system is the essential aspect of the problem. 
Experimentally, small number of carriers in insulators
and semiconductors  introduced by doping or excited by light are key players 
in many important phenomena, where the transport properties of
these carriers are influenced by the polaron effects\cite{Ortmann09} . This is also the case
of doped Mott insulators and the role of electron-phonon coupling is seriously 
considered there in relation to high temperature superconductivity \cite{Mis09}.

The exact solution of the polaron problem is prevented by the highly nonlinear 
and quantum nature of the problem, and most of the analysis is based on 
approximate methods, such as perturbation theory, variational method,
or exact diagonalization of finite size systems. 
It is only recently that the numerically exact solutions for the ground state energy
and optical conductivity at zero temperature have been obtained by
diagrammatic Monte Carlo simulations \cite{MPSS,OC_Froh,OC_Froh1,OC_tJ,OC_Hol,DobCloud}, 
which revealed  pros and cons of various approximate schemes.  
Finite temperature properties are even more difficult to analyze due to 
several technical problems, and more than six decades of efforts to 
understand the temperature dependence of polaron mobility $\mu_{\lambda}(\omega,T)$, 
ranging from already historic papers 
\cite{Holstein59,FrieHol53,Glarum63,LanKad64,Friedman63,Friedman64,
Friedman65,Langreth67,FHIP} to the modern studies 
\cite{Giuggioli03,Cheng03,Troisi06,Chang07,Ciuchi09,Ortmann09,Ortmann10,CaFiPe11}, 
established different behaviors which, however, do not exhaust the regimes listed in the abstract. 

To discuss them we consider the one-dimensional Holstein polaron 
model \cite{Holstein59}:
\begin{equation}
\nonumber {\cal H} = \sum_{k} \left(
\varepsilon^{}_{k}c^{\dagger}_{k}c^{}_{k} + \omega_0
b^{\dagger}_{k} b^{}_{k} \right) + \frac{g}{\sqrt{N}}
\sum_{k, q} c^{\dagger}_{k - q}c^{}_{k}
\left( b^{\dagger}_{q} + b^{}_{-q} \right).
\label{Holstein}
\end{equation}
Here, $c_{k}^{\dagger}$ and $b_{k}^{\dagger}$ are electron and
phonon creation operators in the state of momentum $k$. 
The expression for dispersion $\varepsilon_{\mb{k}} = -2t \cos(k a)$ 
stems from a nearest-neighbor hopping on a linear lattice with lattice
constant $a$, and the Einstein optical phonons have energy
$\omega_0$. The last term describes the local 
electron-phonon coupling (EPC). The units are such that $\hbar=1$.
All sums over momenta are over the Brillouin
zone, and we take the total number of sites $N \to \infty$. 
The charge current operator of the model is $\hat{j}= 2 e a t \sum_{k} \sin(k a) 
c_{k}^{\dagger}c_{k}^{}$, where $e$ is the electron charge.
The strength of the EPC is measured by 
the dimensionless coupling constant $\lambda=g^2 / (2t\omega_0)$,
which sets the borderline between weak- and strong-coupling regime
at $\lambda_c$ of the order of unity. 
For one polaron we introduce the dynamic mobility 
$\mu_{\lambda}(\omega,T) = \sigma_{\lambda}(\omega,T)/e$ as a quantity related 
to the optical conductivity (OC) $\sigma_{\lambda}(\omega,T)$.
The static mobility $\mu_{\lambda}(T) = \mu_{\lambda}(\omega \to 0,T)$ is just the 
direct current (d.c.) characteristics measured in standard transport experiments.

\begin{figure}
\begin{center}
\includegraphics[width=8cm]{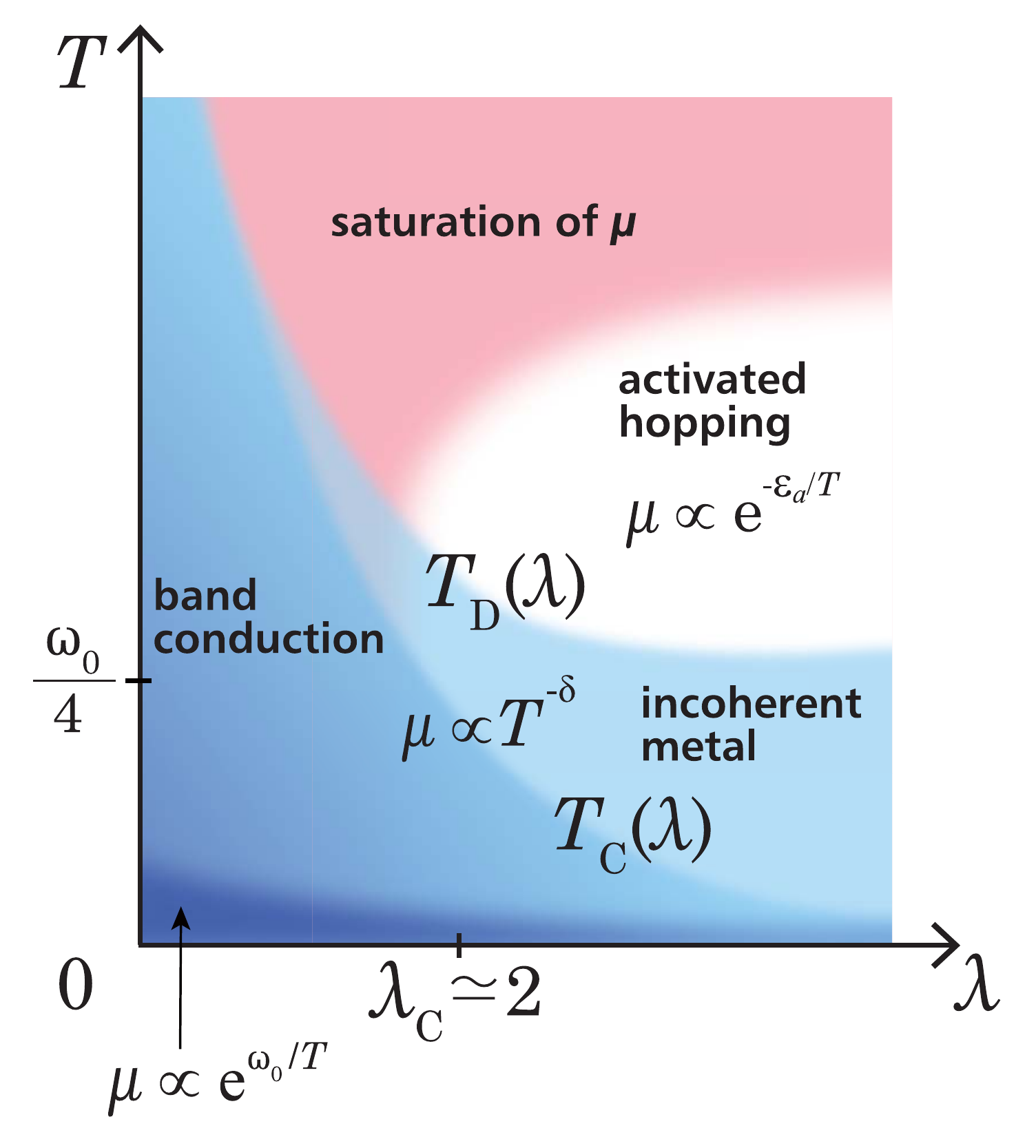}
\end{center}
\caption{\label{fig:fig0} 
Transport regimes of polaron. 
Schematic phase diagram showing the four different regimes of polaron mobility $\mu$
 in the plane of $\lambda-T$ ($\lambda$: electron-phonon coupling strength, $T$: temperature). 
Here the unit of energy is $t=1$ ($t$: transfer energy) and $k_B=\hbar=1$. $\omega_0$ is 
the phonon frequency (see text for a description of the different transport regions). 
} 
\end{figure}

In Fig.~\ref{fig:fig0} the different transport regions, emerging in the $\lambda-T$ plane, 
are highlighted.
The regime of  {\it exponentially large mobility} is well known from early 
analytic studies\cite{Langreth67}. 
At very low $T$ the thermal population of phonons is reduced as $\sim e^{ - \omega_0/T}$ 
and $\mu_{\lambda}(T \to 0) \sim  e^{ \omega_0/T}$ is expected in the lowest 
temperature region, although our numerical studies could not reach such region.
As it was revealed from our numeric results,
there are two main crossovers separating distinct regions at higher temperatures.
One is the temperature $T_D(\lambda)$ above which the Drude peak in 
$\mu_{\lambda}(\omega,T)$ disappears. The second is the temperature $T_C(\lambda)$ above which
the mean free path (MFP) $\ell_{\mbox{\scriptsize MFP}}$ becomes shorter than the lattice constant. 
We found that $T_D(\lambda)$ coincides with the change in the temperature dependence 
of $\mu_{\lambda}(T)$, while $T_C(\lambda)$ indicates the crossover from the
{\it band conduction} to the {\it incoherent metallic} motion.
Namely, in spite of an always metallic $T$-dependence $d \mu_{\lambda}(T) /dT<0$  
at $T<T_D(\lambda)$, one cannot always assume that the description in terms 
of a standard band motion  with large
MFP is valid. 
Indeed, the band motion takes place only in the left lower corner  of $\lambda-T$ diagram, 
whereas the incoherent metallic
behavior with short mean free path is realized at $T>T_C(\lambda)$.  
Note the absolutely different nature of transport in these two regimes which both demonstrate 
power-law $T$-dependence    
\begin{equation}
\mu_{\mbox{\scriptsize metal}}(T) \sim T^{- \delta } \; 
\label{metal}
\end{equation}  
with the index $\delta \approx 2$ at weak- $\lambda \ll 1$ and $\delta \approx 3$ at
intermediate- and strong-couplings $\lambda \ge 0.5$.

As raising $T>T_D(\lambda)$ the temperature dependence of mobility 
$\mu_{\lambda}(T)$ considerably changes. 
At high temperatures  {\it  mobility saturation} is observed: the steepness of the mobility
temperature dependence, $d \mu_{\lambda}(T) /dT$, becomes considerably smaller
for weak EPC $\lambda< \lambda_c$.
In contrast, at larger EPC, $\lambda >\lambda_c$, and lower temperatures 
(but still at $T>T_D(\lambda)$), a different transport regime,  the well known {\it activated hopping}, sets in: $d \mu_{\lambda}(T) / d T$ 
becomes positive. 
It has been derived analytically
 \cite{FrieHol53,Holstein59,Friedman64,Emin91,Chang07}
and confirmed for specific parameters by our numeric results that in the last case 
\begin{equation}
\mu_{\mbox{\scriptsize hop}}(T) \sim T^{-\kappa} \exp(-\varepsilon_a/T) \; ,
\label{hopping}
\end{equation}  
where $\kappa=1$ ($\kappa=3/2$) for adiabatic $\omega_0 \ll t$
(nonadiabatic $\omega_0 \gg t$) case. 
Here, the activation energy  $\varepsilon_a$ is
\begin{equation}
\varepsilon_a = E_b / 2 - t' \; ,
\label{hop}
\end{equation}  
where $E_b$ is a polaron binding energy and $t'=t$
($t'=0$) in the adiabatic (antiadiabatic) case.
The hopping transport begins above a temperature which has been derived to lie  
in the range between $\omega_0/4$ and $\omega_0/2 $  \cite{Holstein59,Friedman64}. 
As we found in our studies, at high enough temperatures, the mobility $\mu_{\lambda}(T)$ 
tends to saturate also at strong EPC. 
Moreover,  at all couplings, weak or strong, the mobilities  converge to 
close values which are almost  independent of EPC $\lambda$, at least in the logarithmic scale.
Our unbiased method giving the basis for the phase diagram shown in
 Fig.~\ref{fig:fig0} is presented in online supplementary information.

Now we present numeric data for the model 
(\ref{Holstein}) at $t=1$, $\omega_0=t$. Also $a$, the Boltzmann constant $k_B$ 
and $e$ are set to unity throughout the paper.
The goal is to get $\mu_{\lambda}(\omega,T)$ in the weak, intermediate, and strong EPC in 
a wide range of temperatures. 
Crossover from weak $\lambda \ll 1$ to strong $\lambda \gg 1$ regime is extremely
smooth in one dimension \cite{Bonca02,Kornil06}. 
So, we used Diagrammatic Monte Carlo method \cite{MPSS} to find the value of coupling 
$\lambda_c$ constant dividing these regimes and calculated the effective 
mass renormalization
 $m^{*}/m_0$, the binding energy $E_b$, and the mean number of phonons 
$\langle H_{ph} \rangle$ ($H_{ph}=\sum_{k} b^{\dagger}_{k} b^{}_{k} $) in the
phonon cloud of a single polaron (Fig.~\ref{fig:fig1} (b), (c), and (d)) at $T=0$.
The second derivatives $d^2 \langle H_{ph} \rangle / d \lambda^2$ and
$d^2  (m^{*}/m_0) / d \lambda^2$
change signs at $\lambda_c \approx 2$. Therefore, we define weak ($\lambda<\lambda_c$), intermediate ($\lambda\simeq\lambda_c$) and strong $\lambda>\lambda_c$ coupling regions.  
Also the temperature dependence of the kinetic energy (Fig.~\ref{fig:fig1}(a)) 
suggests just such a value for $\lambda_c$. 
Indeed the average value of the kinetic energy, $\langle -\hat{K_{xx}} \rangle $, 
is monotonic (non-monotonic) function of $T$
at $\lambda<\lambda_c$ ($\lambda>\lambda_c$) with crossover value $\lambda_c \cong 2$.
 
\begin{figure}
\begin{center}
\includegraphics[width=8cm]{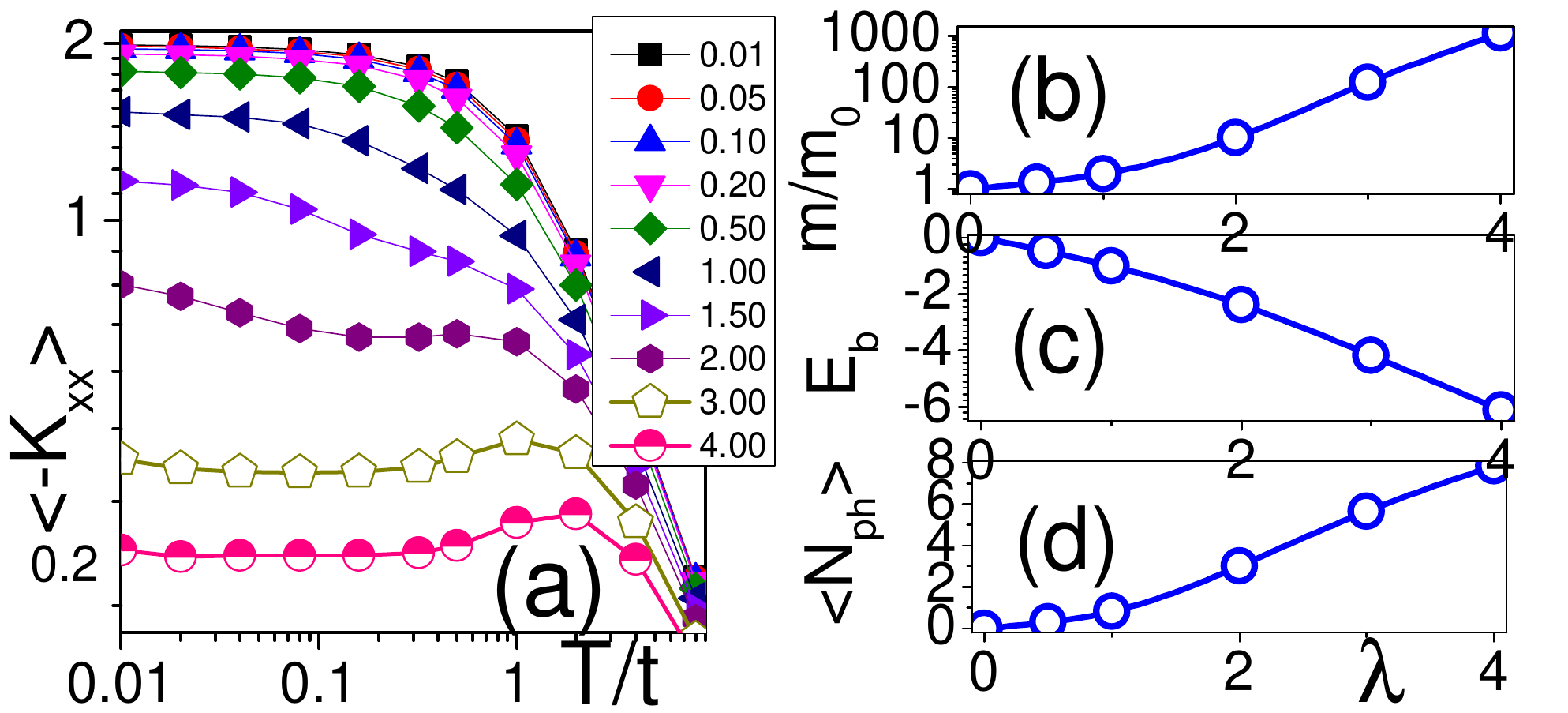}
\end{center}
\caption{\label{fig:fig1} 
Dependence of polaron properties on $\lambda$.
(a) Temperature dependence of the kinetic energy (in units of $t$). 
Dependence at $T=0$ of: (b) 
the effective mass $m^*/m_0$, (c) the binding energy (in units of $t$), and (d) the mean 
number of phonons in the polaron cloud.} 
\end{figure}

\begin{figure}
\begin{center}
\includegraphics[width=8cm]{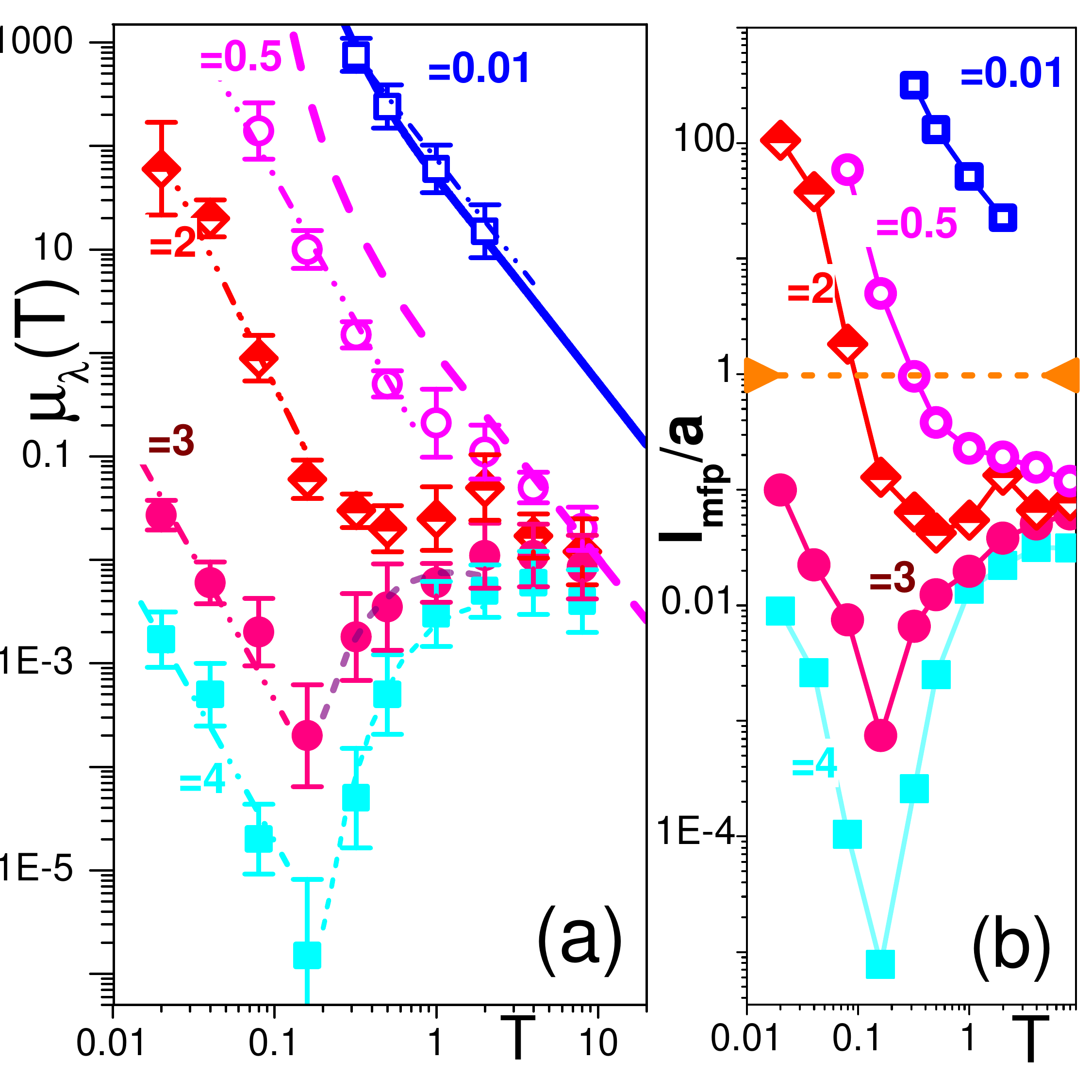}
\end{center}
\caption{\label{fig:fig2} 
Temperature dependence of polaron mobility and mean free path.
(a) D.c. mobility $\mu_{\lambda}(T)$ 
(in units of $e a^2 /\hbar$). 
Unbiased numeric values at $\lambda=0.01$ (open squares), $\lambda=0.5$ (open circles), 
$\lambda=2$ (semi-filled diamonds), $\lambda=3$ (filled circles), and
$\lambda=4$ (filled squares).   
Solid bold ($\lambda=0.01$) and dashed bold ($\lambda=0.5$) lines in the top part 
of figure show the results obtained by the Boltzmann approach \cite{BolApp}.
Fit of the mobility by the activation law Eqs.~(\ref{hopping}) and (\ref{hop}) 
is shown for $T>0.2$ at $\lambda=3$ (short-dash line) and $\lambda=4$ (dotted line). 
Linear dash-dot-dot lines are fits of the low temperature contribution
of mobility, for all the values of $\lambda$, 
by a power law $\mu \sim T^{-\delta }$; (b) mean free path, in units of the 
lattice parameter $a$, vs temperature (the symbols 
are the same as those used in panel (a).
} 
\end{figure}

In Fig.~\ref{fig:fig2}a we present $\mu_{\lambda}(T)$ in the perturbative ($\lambda=0.01$), 
weak ($\lambda=0.5$), intermediate ($\lambda=2$), and strong 
($\lambda=3$ and $\lambda=4$) 
EPC limits. 
First, we show that we reproduce the Boltzmann result in the perturbative limit
(compare open squares and solid bold line in Fig.~\ref{fig:fig2}a).
A perturbative low $\lambda$ analytic treatment of the Holstein model predicts power laws 
(\ref{metal}) with $\delta=3/2$ for $T \ll t$ and $\delta=2$ for $T \gg t$ \cite{Glarum63}. 
Our data (dash-dot-dot line fitting positions of the open squares in 
Fig.~\ref{fig:fig2}a) in the range $0.32<T<2$ support the value $\delta=2$. 
We restricted our analysis  to $T> 0.3$ at $\lambda=0.01$  
because of the instability of the spectral analysis for extremely narrow ($<10^{-3}$) and 
high ($>10^{3}$) Drude peaks (see  online supplementary information).       

\begin{figure}
\begin{center}
\includegraphics[width=7cm]{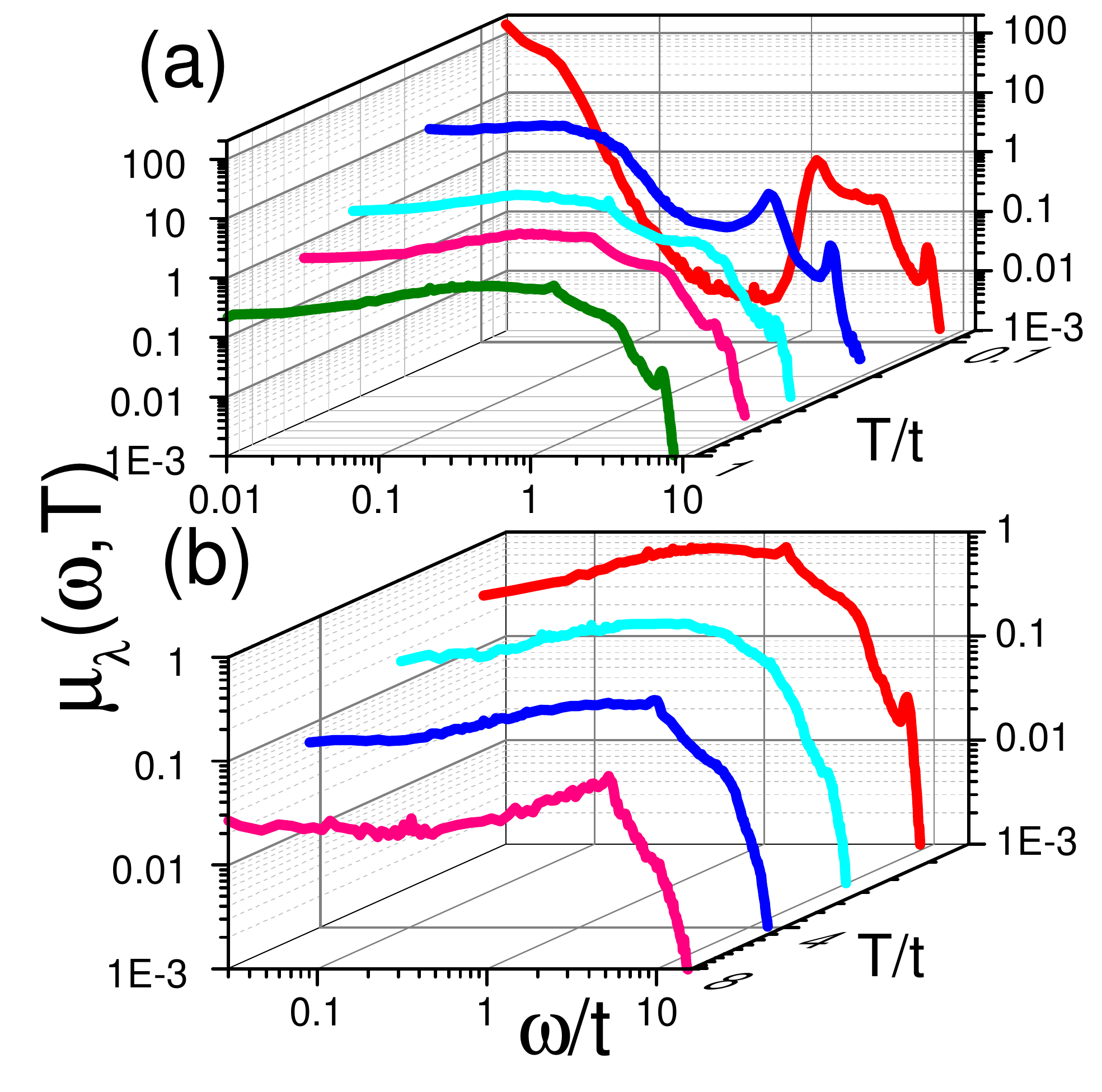}
\end{center}
\caption{\label{fig:fig3}  Dynamic mobility in the weak 
coupling regime (in units $ea^2/\hbar$): 
$\lambda=0.5$ at (a) low  ($T=0.08, 0.16, 0.32, 0.5, 1$) and (b) high ($T=1, 2, 4, 8$) temperatures.
} 
\end{figure}

The Boltzmann (bold dashed line) and unbiased (open circles) results are already different 
at $\lambda=0.5$ (Fig.~\ref{fig:fig2}a).  Actually, for larger couplings, $\lambda \ge 0.5$, 
one can always fit $\mu_{\lambda}(T)$ by a power law (\ref{metal}) below some $\lambda$-dependent
 temperature $T_{D}(\lambda)$: $T_{D}(\lambda=0.5) \approx 0.5$
and $T_{D}(\lambda \ge 2) \approx 0.25$. 
However, the index of the power law is different from $\delta=2$ 
found in the perturbative $\lambda \ll 1$ limit. 
The best fit provides  $\delta \approx 2.8$ for $\lambda=0.5, 3, 4$ and 
$\delta \approx 3.2$ for $\lambda=2$ (dash-dot-dot lines
fitting low temperature positions of open circles, diamonds, filled circles, and
filled squares in Fig.~\ref{fig:fig2}a).
Our result is consistent with an exponent $3/2 \lesssim \delta \lesssim 3$ which is 
experimentally observed in many different materials
\cite{Giuggioli03,Cheng03,Kao81,Jurchesku04,Dimi01,Klauk99,Warta85,Nelson98,Karl01,Gers06}.
We emphasize that the $d\mu_{\lambda}(T)/dT<0$ behavior at low temperatures cannot be 
regarded as a proof of weak EPC.  

For $T>T_D(\lambda)$ one can observe mobility saturation at $\lambda\leq 2$ whereas, at $\lambda=3,4$, hopping transport followed by mobility saturation.
In particular, the hopping transport  is naturally distinguished from the resistivity saturation 
regime by the existence of temperature range with positive derivative   
$d \mu_{\lambda}(T) / d T>0$ above a characteristic temperature 
whose analytical estimate is in the range  between $\omega_0/4$ and $\omega_0/2$
\cite{Holstein59,Friedman64}. 
We get a value consistent with $\approx \omega_0 / 4$ 
(filled circles and filled squares for mobilities at $\lambda=3$ and $\lambda=4$ in Fig.~\ref{fig:fig2}a).
The analytic value of the activation energy $\varepsilon_a$ of the activation law (\ref{hopping})
is related to the binding energy of polaron $E_b$ in Eq.~(\ref{hop}). 
Inserting the binding energy of the polaron $E_b=4.19$ ($E_b=6.14$) at 
$\lambda=3$ ($\lambda=4$) into Eq.~(\ref{hop}), one obtains $\varepsilon_a=1.1$
($\varepsilon_a=2.07$) which is very close to the value 1.2 (2.1) obtained by the fit
of $\mu_{\lambda}(T>0.3)$ at $\lambda=3$ ($\lambda=4$) (see short-dash
(dotted) line fitting high temperature dependence of filled circles (squares) in Fig.~\ref{fig:fig2}a).
Note, the fit is consistent only with the estimate of the activation energy in Eq.~(\ref{hop}) 
corresponding to the adiabatic regime.   

So far, our analysis can distinguish a low-$T$ regime at $T<T_D(\lambda)$, where power law 
decrease of mobility is observed, and, at $T>T_D(\lambda)$, two different regimes depending on temperature and EPC.
However, it is clear that low-$T$ regimes must be different at $\lambda \ll 1$ and 
$\lambda \gg 1$ because increase of $\lambda$ must eventually encounter Mott-Ioffe-Regel 
limit for mean free path (MFP) $\ell_{\mbox{\scriptsize MFP}}$ where band conduction with 
$\ell_{\mbox{\scriptsize MFP}}>a$ changes 
to an incoherent metallic transport with $\ell_{\mbox{\scriptsize MFP}}<a$.
To estimate the mean free path $\ell_{\mbox{\scriptsize MFP}}$ we write the optical absorption 
$\sigma(\omega)=-i N_e \langle \hat{K}_{xx} \rangle /(\omega+iM(\omega))$
in terms of the memory function $M(\omega)$ \cite{mori65,mori65a}.
At $\omega=0$ the function $M$ is real and determines the reciprocal of the 
optical relaxation time $1/\tau_r$, so that the mobility turns to be 
$\mu=- \langle \hat{K}_{xx} \rangle \tau_r$. 
This last relation allows to extract $\tau_r$. The free mean path is 
defined by $\ell_{\mbox{\scriptsize MFP}}=v \tau_r$, and a rough estimate of the average velocity 
$v$ can be obtained by
$v \simeq \sqrt{ \langle \hat{j}(0) \hat{j}(0) \rangle }$. In Fig.~\ref{fig:fig2}b we plot the temperature dependence of 
$\ell_{MFP}$ at the different values of $\lambda$.

As shown above, analysis of our data distinguish four regimes. 
Two low-$T$ regimes, band conduction and incoherent metallic transport, 
are characterized by the power law decrease of mobility when $T$ increases. 
These two regimes are distinguished by the MFP, which is much larger than the lattice constant 
in the first case and much smaller than $a$ in the second case.  
They are separated from the two high-$T$ regimes by $\lambda$-dependent 
temperature $T_D(\lambda)$, where metallic temperature dependence of mobility significantly changes. 
It becomes slower in high temperature saturation regime which sets up above $T_D(\lambda)$
at $\lambda\le\lambda_c$.
To the contrary, $\mu_{\lambda}(T)$ starts to increase with temperature at large EPC 
$\lambda > \lambda_c$, although it also eventually saturates at large $T$.  
Careful analysis of the frequency dependent mobility $\mu_{\lambda}(\omega,T)$ 
shows that $T_D(\lambda)$ separates profoundly different physical regimes. 
Namely, the Drude peak is observed only below $T_D(\lambda)$.   
We note that the temperature $T_D(\lambda)$, where 
the Drude peak disappears, coincides with
temperature where conductivity saturation starts or  with the  temperature 
where the activated hopping regime arises  \cite{Holstein59,Friedman64}.

\begin{figure}
\begin{center}
\includegraphics[width=8cm]{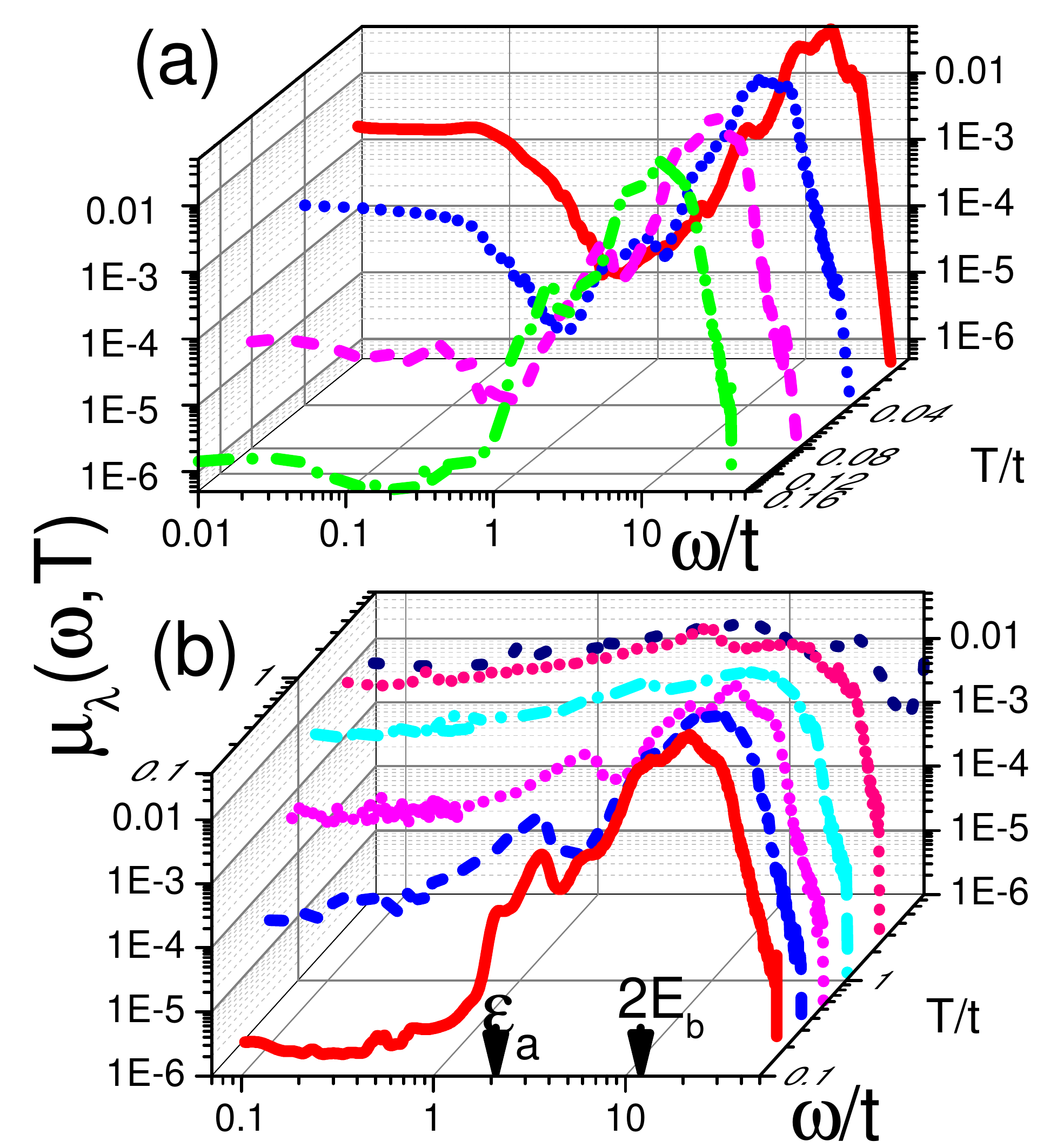}
\end{center}
\caption{\label{fig:fig4} Dynamic mobility in the strong coupling regime (in units $ea^2/\hbar$):
$\lambda=4$ at (a) low  
($T=0.02, 0.04, 0.08, 0.16$) and (b) high
($T=0.16. 0.32, 0.5, 1, 2, 8$) temperatures. 
Note opposite direction of T-axis in (a) and (b). 
Arrows in the bottom of the figure show the activation 
energy $\varepsilon_a$ from Eq.~(\ref{hop}) and 
twice the polaron binding energy $2E_b$, respectively.
} 
\end{figure}

Figure~\ref{fig:fig3} shows temperature dependence of the $\mu_{\lambda}(\omega,T)$ in 
the weak EPC, $\lambda=0.5$. 
Low-energy Drude peak is clearly seen at low $T$, $T \le 0.16$, it almost 
vanishes at $0.32<T<0.5$, and it is absent for $T>0.5$. 
The mobility in the band conduction regime quickly decreases with temperature 
at  $T<0.32$, which coincides with the temperature range of the Drude peak existence.
For higher temperatures, in agreement with assumptions made in
 Refs.~\cite{GuCaHa,GuHa,CaGu01,CaGu02}, the resistivity saturation occurs.
Furthermore, we found that  the Drude peak at $\lambda=0.5$ and $T<0.32$ 
gradually disappears without significant change  of high energy-part, which
is again in complete agreement with Refs.~\cite{GuCaHa,GuHa,CaGu01,CaGu02}.  

The $T$-dependence of the $\mu_{\lambda}(\omega,T)$ at strong EPC, 
$\lambda=4$ (Fig.~\ref{fig:fig4}), also supports the statement that the large negative derivative
$d\mu_{\lambda}(T)/dT<0$ is associated with the presence of a Drude peak.
Indeed, $d\mu_{\lambda}(T)/dT<0$ at $T \le 0.16$ (Fig.~\ref{fig:fig2}) which is just the range 
where Drude peak is seen (Fig.~\ref{fig:fig4}(a)).  As previously discussed, this regime is not related
to a band conduction transport, but it stems from a incoherent motion
of the charge carriers with short mean free path. 
To the contrary, a Drude-like peak 
is absent (Fig.~\ref{fig:fig4}(b)) in the domain of the thermally activated 
transport, $T \ge 0.25$ (Fig.~\ref{fig:fig2}).   
It is known \cite{Emin93} and confirmed in our study that the OC is 
characterized by a broad peak with the 
maximum around twice of the binding energy (Fig.~\ref{fig:fig4}b). 
At $0.25<T<2$ we find that the $\mu_{\lambda}(\omega,T)$ is $T$-independent at
 $\omega>\varepsilon_a$ 
while the spectral weight at $\omega<\varepsilon_a$ growth exponentially when $T$ increases
(Fig.~\ref{fig:fig4}b). 
Then, the spectral weight starts to spread to larger frequencies when $T>2$ and, as a result, the 
static mobility $\mu_{\lambda}(T)$ starts to saturate.

In conclusion, we presented for the first time unbiased results for the temperature dependence of 
the optical conductivity $\sigma_{\lambda}(\omega,T)$ (or dynamic mobility 
$\mu_{\lambda}(\omega,T)$) and static mobility $\mu_{\lambda}(T)$ of the 
one-dimensional Holstein polaron. 
The transport features display a strong $\lambda$ and $T$ dependence. In particular we 
proved that a low-$T$ power-law behavior exists
until the Drude peak in the OC disappears at $T_D(\lambda)$. However, while
the standard band-like transport is recovered at weak couplings, an unconventional incoherent
regime is observed at larger couplings. 
Moreover, at $T>T_D(\lambda)$, the $\mu$-saturation (activated hopping transport) 
phenomenon occurs at weak (strong) couplings. 
Finally our data imply that although mobilities and mean free paths at different values 
of $\lambda$ differ by many orders of magnitude at small temperatures, their values at 
$T>4t$ become very close to each other (Fig.~\ref{fig:fig2}).
Namely, regardless of the strength of the EPC, the effective 
scattering of a polaron turns to be very strong when
the temperature exceeds the bare bandwidth $4t$.

N.N. is supported by Grant-in-Aids for Scientific Research (S) (No. 24224009) from the Ministry of 
Education, Culture, Sports, Science and Technology (MEXT) of Japan and Strategic International 
Cooperative Program (Joint Research Type) from Japan Science and Technology Agency.

\section{Supporting online material}

\subsection{Method to obtain mobiliy}

Beyond the analytic and semi-analytic approaches giving the functional dependencies 
at the various limiting cases 
\cite{Holstein59,FrieHol53,Glarum63,LanKad64,Friedman63,Friedman64,Friedman65,Langreth67,
Giuggioli03,Cheng03,Chang07,Ortmann09,Ortmann10}
there is also a plenty of numeric methods. 
Mobility is calculated adopting variational approaches \cite{FHIP},   
using frozen lattice approximation \cite{Ciuchi09}, through the Einstein relation 
from the diffusion coefficient computed in the semiclassical phonon 
approximation \cite{Troisi06}, and using Dinamical Mean Field 
Theory \cite{Millis99,Fratini03}.
It was also suggested to obtain the mobility from the low frequency limit  
$
\mu = (1/ e) \sigma(\omega \to 0^+) 
$
where the OC $\sigma(\omega)$ is calculated by quantum or semiclassical approach on a rather small finite systems
\cite{CaFiPe11,GuCaHa,GuHa,CaGu01,CaGu02}. 
Note that the finite size of the system, especially in the semiclassical limit \cite{CaFiPe11}, 
implies zero mobility unless {\it arbitrary} artificial level broadenings are added.  
Note, each of the above approaches includes uncontrollable approximation. 

In contrast, all the approximations can be avoided if $\mu$ is obtained from 
the low frequency limit of the OC and $ \sigma(\omega)$ is calculated exactly in the 
infinite lattice system.  
Then, the exact OC $\sigma(\omega)$ can be obtained from the {\it analytic continuation},
i.e. from the solution of the integral equation (see, e.g.~\cite{GuHaSa10})
\begin{equation}
\Pi(\tau) = \int_{-\infty}^{\infty} d\omega \; \frac{1}{\pi} 
\frac{\omega \exp[-\tau \omega]}{1-\exp[-\beta\omega]} \;  \sigma(\omega)
\equiv {\cal F}[\tau,\sigma(\omega)] \; 
\label{s1}
\end{equation}
($\beta=1/T$) if the current-current correlation function 
$
\Pi(\tau) = \langle \hat{j}(\tau) \hat{j}(0) \rangle \;    
$
is exactly known at imaginary times $\tau$. 
Nowadays, unbiased $\Pi(\tau)$ can be obtained for infinite systems 
\cite{OC_Froh,OC_Froh1,OC_tJ,OC_Hol} 
by several exact methods. 
In particular, we used  
the Diagrammatic \cite{MPSS} and World-Line Monte Carlo \cite{Candia}
 methods checking that both approaches give the same results.  
So, the last obstacle to get the exact mobility is to solve the equation (\ref{s1}).  
Unfortunately, it belongs to the class of ill posed problems, i.e., also due 
to the noise, even small, present in 
$\Pi(\tau)$, there is not an unique
function $\widetilde{\sigma}(\omega)$ which 
exactly satisfies $\Pi(\tau_i)-{\cal F}[\tau_i,\widetilde{\sigma}(\omega)] =0$ for all points 
$\tau_i$, $i=1,M$, where $\Pi(\tau)$ is known. 
So, a natural formulation of the strategy to solve the equation (\ref{s1}) is to find a set 
of solutions  
$\widetilde{\sigma}(\omega)$ which minimize the objective function 
\begin{equation}
{\cal O}\left[ \Pi, \widetilde{\sigma} \right] = \frac{1}{M} \sum_{i=1}^M 
\frac{\left\{\Pi(\tau_i)-{\cal F}[\tau_i,\widetilde{\sigma}(\omega) \right\}^2}{s_i} 
\label{s2}
\end{equation}
($s_i$ depend on the method), where $\widetilde{\sigma}(\omega)$ 
is normalized as $\int_{-\infty}^{\infty} d \omega \widetilde{\sigma}(\omega) = \widetilde{N}$
and $\widetilde{N}$ is fixed by the average value 
of the kinetic energy $\hat{K}_{xx} = - 2t \sum_{k=1}^{N} \cos(k) c_k^{\dagger} c_k$
\begin{equation}
\widetilde{N} = -\frac {\pi}{N} \langle \hat{K}_{xx} \rangle 
\; . 
\label{s3}
\end{equation}
Any naive attempt to get $\widetilde{\sigma}$ by a mere minimization
of ${\cal O}$ leads to the saw tooth instability (STI) of the solution $\widetilde{\sigma}(\omega)$ 
when the amplitude of fast oscillations is much larger than the actual answer.
To deal with STI \cite{GuHaSa10,MisJu12} the vast majority of the methods 
introduces a regularization functional ${\cal F}(\widetilde{\sigma})$, 
i.e. ${\cal O} \to {\cal O} + {\cal F}(\widetilde{\sigma})$: 
${\cal F}(\widetilde{\sigma})$ imposes constraints,  
(smoothness, positiveness) on the solution $\widetilde{\sigma}(\omega)$
\cite{Tikh-Arse,JaGu}. 
However, it was shown that such constraints can lead to significant bias of the result 
\cite{Vafayi07} and, hence, it is very important to avoid any disguise of the objective
function by a regularization functional.
In this sense, one can single out sampling approaches \cite{Sand98,MPSS,Vafayi07}
where STI is suppressed by the self-averaging of the noise in a superposition of multiple solutions, 
each having its own large STI. 
Although more computationally involved, these methods are not biased by the 
regularization term. In particular, we used the stochastic optimization method
\cite{MPSS} where $s_i = [\Pi(\tau_i)]^{\kappa}$ ($0<\kappa<1$) 
is a factor highlighting small values of $\Pi(\tau_i)$ when $\Pi(\beta/2)/\Pi(0) \ll 1$.
Description of the approach to determine on the errorbars of the mobility is given in the next section.  

\subsection{Estimation of the errorbars}

There are several sources of error stemming from the estimate 
of the mobility through the low frequency limit of the OC: 
normalization, extrapolation and systematic errors. 
They are related to the properties of the kernel of the integral equation (\ref{s1}),
and to the stochastic optimization method (SOM) of the analytic continuation. 
In particular, the SOM strategy is based solely on the minimization of the objective 
function (\ref{s2}), and does not 
use any regularization strategy to treat the saw-tooth noise instability by distorting the 
objective function. Indeed, SOM averages over many solutions obtained through the minimization of the 
objective function starting from randomly chosen initial conditions. Here 
the first source of uncertainty comes from the normalization of the OC: it
introduces the {\it normalization errorbar}. 
Furthermore, the $\omega \to 0$ limit of the OC requires the extrapolation of the 
OC from low frequencies and sets the {\it extrapolation errorbar}. 
Finally, the {\it systematic errorbar} is related to 
the failure to reproduce exactly the shape of the OC from an ideal noiseless current-current 
correlation function $\Pi(\tau)$.
As a typical example, Fig.~\ref{fig:fig0} shows the three absolute errorbars
as function of the temperature at $\lambda=0.5$ and $\lambda=4$. 
Below we give the description of the procedure to determine the errorbars.   
 
\begin{figure}[thb]
\begin{center}
\includegraphics[width=7cm]{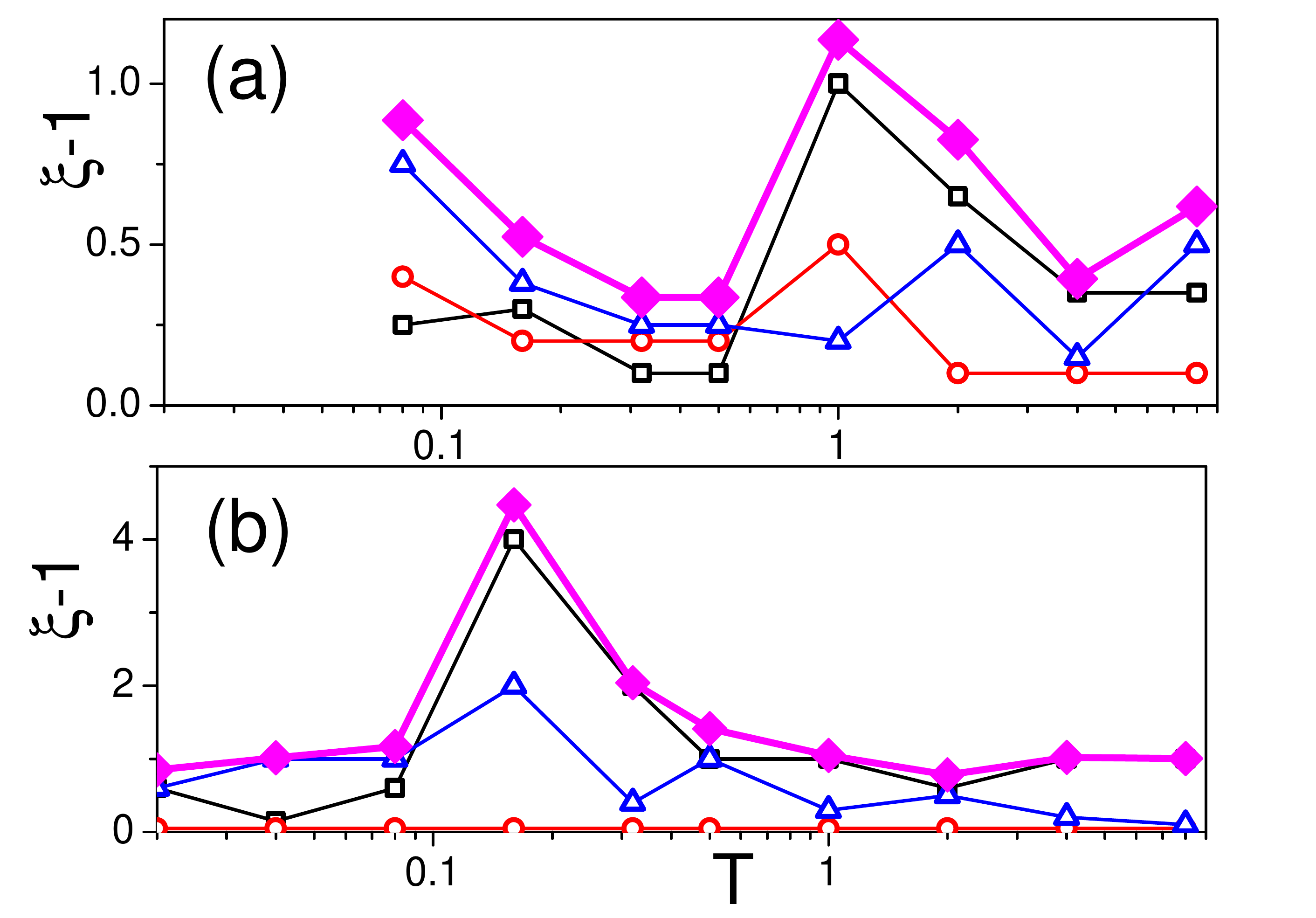}
\end{center}
\caption{\label{fig:fig0} Temperature dependence of the absolute errorbars $\xi-1$ at
(a) $\lambda=0.5$ and (b) $\lambda=4$. Systematic (squares), normalization (circles), 
and extrapolation (triangles) errorbars are summed to obtain the total (diamonds) errorbar.
} 
\end{figure}

The maximal relative errorbar $\xi^{(l)}$ from the given source $l$ 
(l is an index varying between 1 and 3 and corresponds 
to the three different sources of errorbars) is 
obtained in the following way.
We define ${\bar \mu}$ as the true value of the mobility, which is 
either an average over the possible values of $\mu$ depending on some external parameter
(e.g. normalization of OC or range of $\omega$ values used for the extrapolation) or it is
just the exact $\sigma(\omega \to 0)$ value when ideal data are considered.  
The maximal (minimal) possible value of $\mu$, allowed by uncertainty stemming 
from the given error source, is 
$\mu_{\mbox{\scriptsize max}} = \xi_{\mbox{\scriptsize max}}^{(l)} {\bar \mu}$
($\mu_{\mbox{\scriptsize min}} = {\bar \mu} / \xi_{\mbox{\scriptsize min}}^{(l)}$)
Then, the relative errorbar $\xi^{(l)}$, for the assigned source $l$, is defined as the maximum of 
two factors
$\xi^{(l)}=\mbox{MAX}[\xi_{\mbox{\scriptsize max}}^{(l)}, \xi_{\mbox{\scriptsize min}}^{(l)}]$.
We note that the goal of our procedure is to give the upper bound of
the relative errorbar: it avoids the possibility of an its underestimate.

To get the joint influence of the errors from different and independent sources we convert 
the relative errorbars $\xi_l$ into the absolute ones $\delta \xi^{(l)} = \xi^{(l)} -1$. 
It is well known that the absolute errorbars, $\delta \xi_l$, stemming 
from independent sources, can be combined into the total absolute errorbar $\delta \xi =\xi-1$ as  
$\delta \xi = \sqrt{\sum_l [\delta \xi^{(l)}]^2}$. Hence, 
\begin{equation}
\xi = 1 + \sqrt{\sum_l [\xi^{(l)} - 1]^2 } \; . 
\label{s4}
\end{equation}
Our definition insures that the relative errorbar is maximal and implies that the value
of ${\bar \mu}$ is in the range $[{\bar \mu} / \xi , \xi {\bar \mu}]$.  
Note that in logarithmic scale it means
\begin{equation}
\ln[\mu]  = \ln[{\bar \mu}] \pm \ln[\xi] \; . 
\label{s5}
\end{equation}
In linear scale one has 
$\mu(1-\delta\xi^{(-)}) < \mu < \mu(1+\delta\xi^{(+)})$, where 
$\delta\xi^{(+)}=\xi-1$ and $\delta\xi^{(-)}=1-\xi^{-1}$. 

The analysis presented in Fig.~\ref{fig:fig0} shows that the practical estimate of the errorbars 
provides $\xi > 1$ for some values of the parameters $\lambda$ and $T$.
Such large errorbars imply that the uncertainty of the mobility value
is of the same order or even larger than the obtained value ${\bar \mu}$ itself. 
However, the temperature dependence of the mobility, in the range of the parameters 
investigated in the main text,  
involves ten orders of magnitude (see Fig.~3 of the main text) and, thus, even such an uncertainty 
does not make the analysis senseless. 

\begin{figure}[thb]
\begin{center}
\includegraphics[width=7cm]{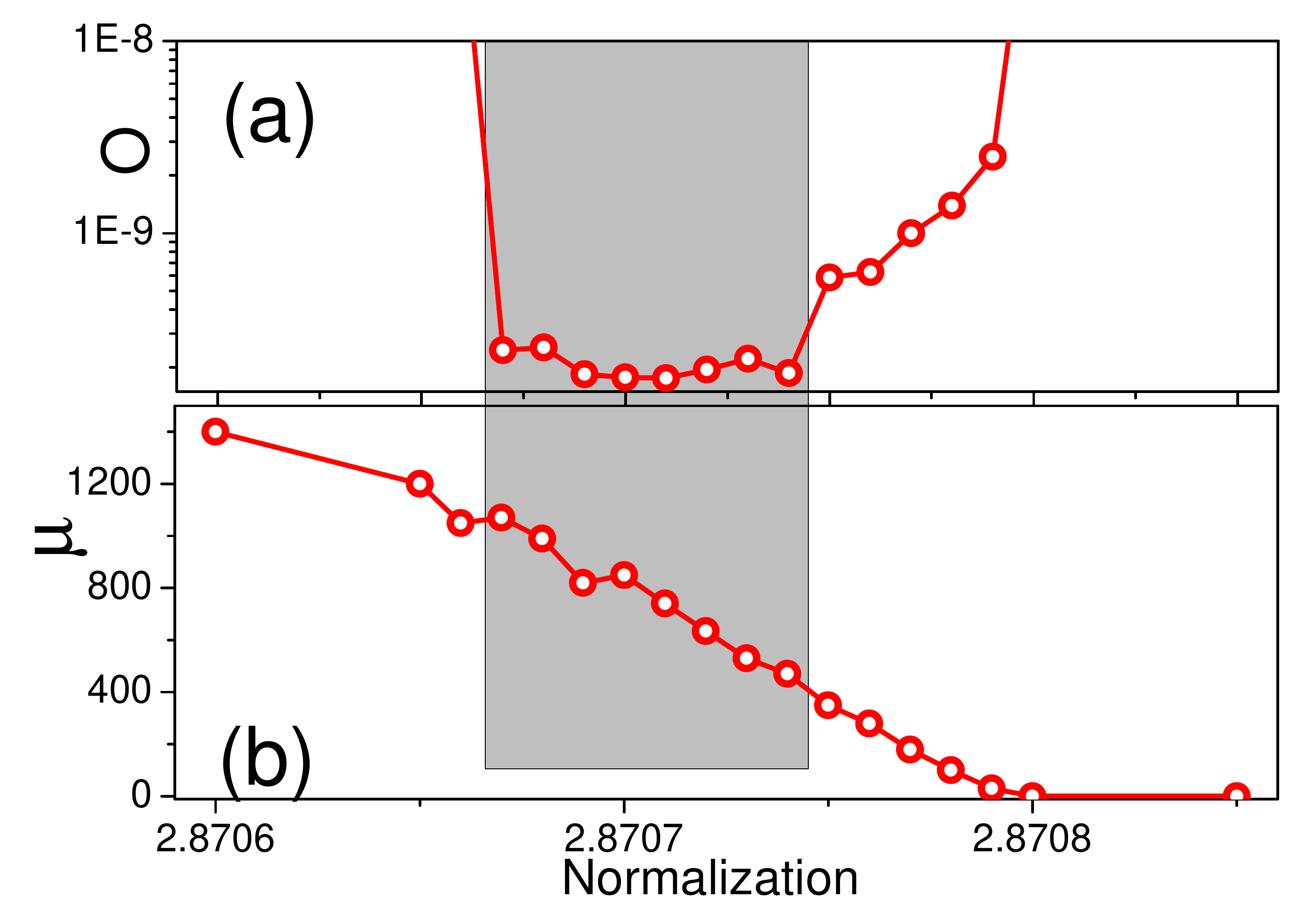}
\end{center}
\caption{\label{fig:fig1} Dependence of (a) the value of the objective function (\ref{s2}) and
(b) the value of mobility on the normalization coefficient $\widetilde{N}$ at
$\lambda=0.01$ and $T=0.32$.} 
\end{figure}

The {\it normalization errorbar} $\xi^{\mbox{\scriptsize {(nor)}}}$ comes
from uncertainty related to the normalization $\widetilde{N}$. 
Naively, it seems that the normalization ${\tilde N}$ of the OC can be fixed by the sum rule 
(\ref{s3}) relating ${\tilde N}$ to the kinetic energy. 
However, the sum rule (\ref{s3}) does not fix the problem since the shape of the OC is very sensitive to 
${\tilde N}$ even in presence of very small errorbars, ${\tilde N}(1 \pm 10^{-4})$, which are 
set by the statistical fluctuations of the average kinetic energy calculated through Monte Carlo 
methods.     

\begin{figure}[thb]
\begin{center}
\includegraphics[width=8cm]{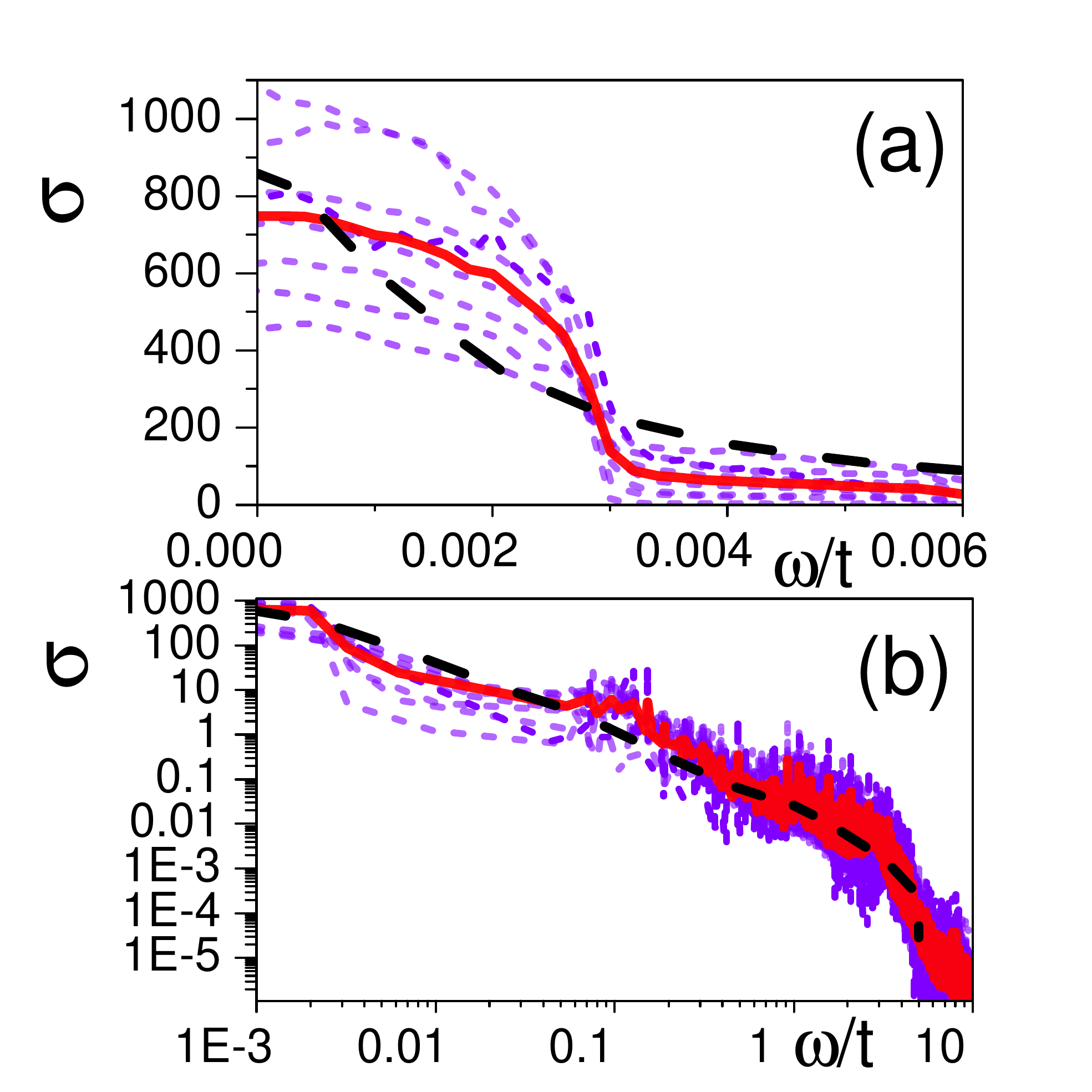}
\end{center}
\caption{\label{fig:fig2} OC for different values of the normalization coefficient 
$\widetilde{N}$ (short-dashed
lines with $\widetilde{N}$ increasing from top to bottom) (a) in the low frequency part 
and (b) the whole range in logarithmic scale. Solid line is the 
average OC in the range of adopted normalizations and 
long-dashed line is the Boltzmann result  in the perturbative limit.} 
\end{figure}

However, the sum rule gives a crude value of the correct normalization and one can notice that
the value of the objective function increases when ${\tilde N}$ is too large or too small
(see Fig.~\ref{fig:fig1}a). 
Then, one can try to find a proper $\widetilde{N}$ by choosing the normalization that provides the 
minimum of the objective function. 
However, this procedure does not give an unique value of the proper 
normalization ${\tilde N}$ because 
there is a range of normalizations where objective function does not change substantially being 
the minimum very flat (Fig.~\ref{fig:fig1}a).
On the other hand OC (Fig.~\ref{fig:fig2}) and mobility $\mu \sim \sigma(\omega \to 0)$ 
(Fig.~\ref{fig:fig1}b) change considerably.  
Figure \ref{fig:fig1} shows the result of this procedure at $\lambda=0.01$ and $T=0.32$. 
The values of mobility $\mu({\tilde N})$ (Fig.~\ref{fig:fig1}b), considered as proper values, 
are restricted to the shaded area which is determined from the flat minimum of the objective function 
$O({\tilde N})$ (Fig.~\ref{fig:fig1}a). Next the average value of the proper mobilities 
is considered (${\bar \mu}$) and the errorbar 
$\delta \xi_{\mbox{\scriptsize nor}}$ is taken as {\it maximum}
deviation from ${\bar \mu}$ in the shaded area. Also $\sigma(\omega)$ (Fig.~\ref{fig:fig2})
is found as the average of the OC values at the different proper ${\tilde N}$. 
We note that this source of error gives a significant contribution at small values of $\lambda$, 
when the current-current correlation function $\Pi(\tau)$ is a very flat function of $\tau$. On 
the other hand it provides an almost negligible contribution 
at large values of $\lambda$ (compare (a) and (b) in Fig.~\ref{fig:fig0}).    
In particular, due to the normalization error,
the analytic continuation at $\lambda=0.01$ is one of the most difficult cases 
among the other ones encountered in the present study.
The reason is that $\Pi(\tau)$ is an almost flat function of the imaginary time.
For example, $\Pi(\beta/2)/\Pi(0) \approx 0.98$ at $T=0.32$ and only the high sensitivity of 
the proposed analytic continuation method can discern a Drude peak with width less 
than $10^{-2}$ and height more than $10^{2}$. Note, the method fails for higher and narrower 
peaks preventing us from the study of the regime where mobility becomes exponentially large 
(very low temperatures).  

\begin{figure}[thb]
\begin{center}
\includegraphics[width=8cm]{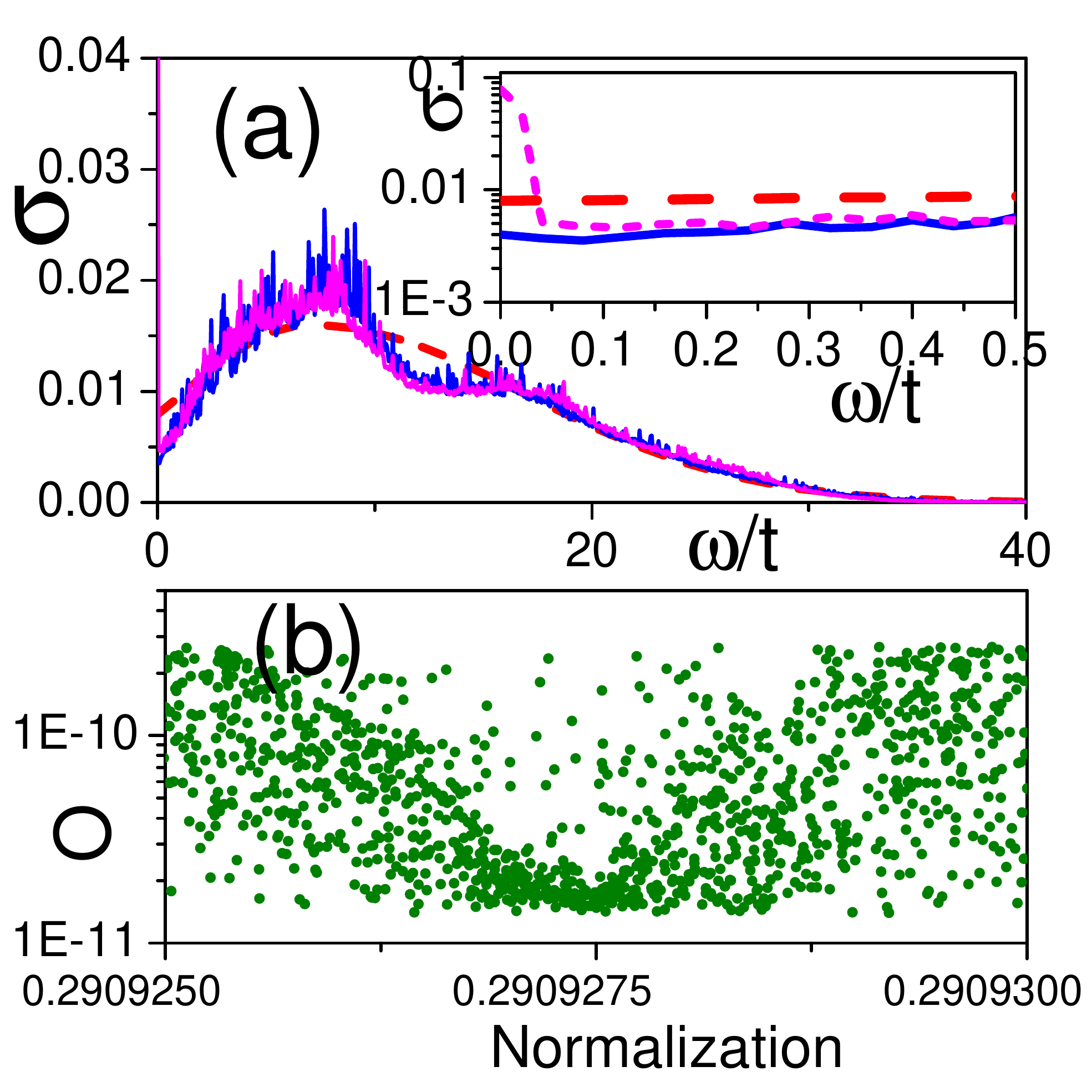}
\end{center}
\caption{\label{fig:fig3} Numeric test for reproducibility of
OC at $\lambda=3$ and $T=8$. (a) "True spectrum" (dashed line)
is compared with the analytic continuation results averaged over 
different ranges of normalization.
(b) Objective functions found through spectral analysis at different normalizations.} 
\end{figure}

The {\it extrapolation errorbar}  $\delta \xi^{\mbox{\scriptsize {(ext)}}}$
comes from the instability of the low frequency behavior of the 
solution $\widetilde{\sigma}(\omega)$.
An example of such an instability is given in Fig.~\ref{fig:fig3}. 
Panel (b) shows the values of the 
objective function obtained by minimizing $O$ at different 
normalizations. 
The function $\Pi(\tau)$ for this analysis was calculated from the model curve for 
$\sigma(\omega)$, called from now on as "true spectrum", and the SOM was used to recover 
the OC from $\Pi(\tau)$.
Panel (a) shows the comparison of the true spectrum with the average of the OCs, obtained by varying 
the normalization coefficient below and above the value ${\tilde N}=0.2909275$.   
We note that the overall shape of the OC is stable though the behavior of the low frequency part
(Inset in Fig.~\ref{fig:fig3}a) is very sensitive to the normalization. 
This kind of instability comes from the structure of the kernel of the integral equation (\ref{s1}): 
it is observed even in the case of noiseless ideal data. 
We found that low-frequency instability is typical at large temperatures, $T \ge 1$, where 
the imaginary time base for the spectral analysis, $\beta=1/T$, is small. 
This helps to identify this kind of instability because the existence of sharp features in the 
frequency range $\Delta \omega \ll T$ is not meaningful from the physical point of view.         
In the case of such an instability the value of the mobility is obtained from the extrapolation of 
the $\widetilde{\sigma}(\omega)$ from finite frequencies.
Naturally, different 
ranges of $\omega$ used for extrapolation can lead to different results for the mobility $\mu$. 

\begin{figure}[thb]
\begin{center}
\includegraphics[width=8cm]{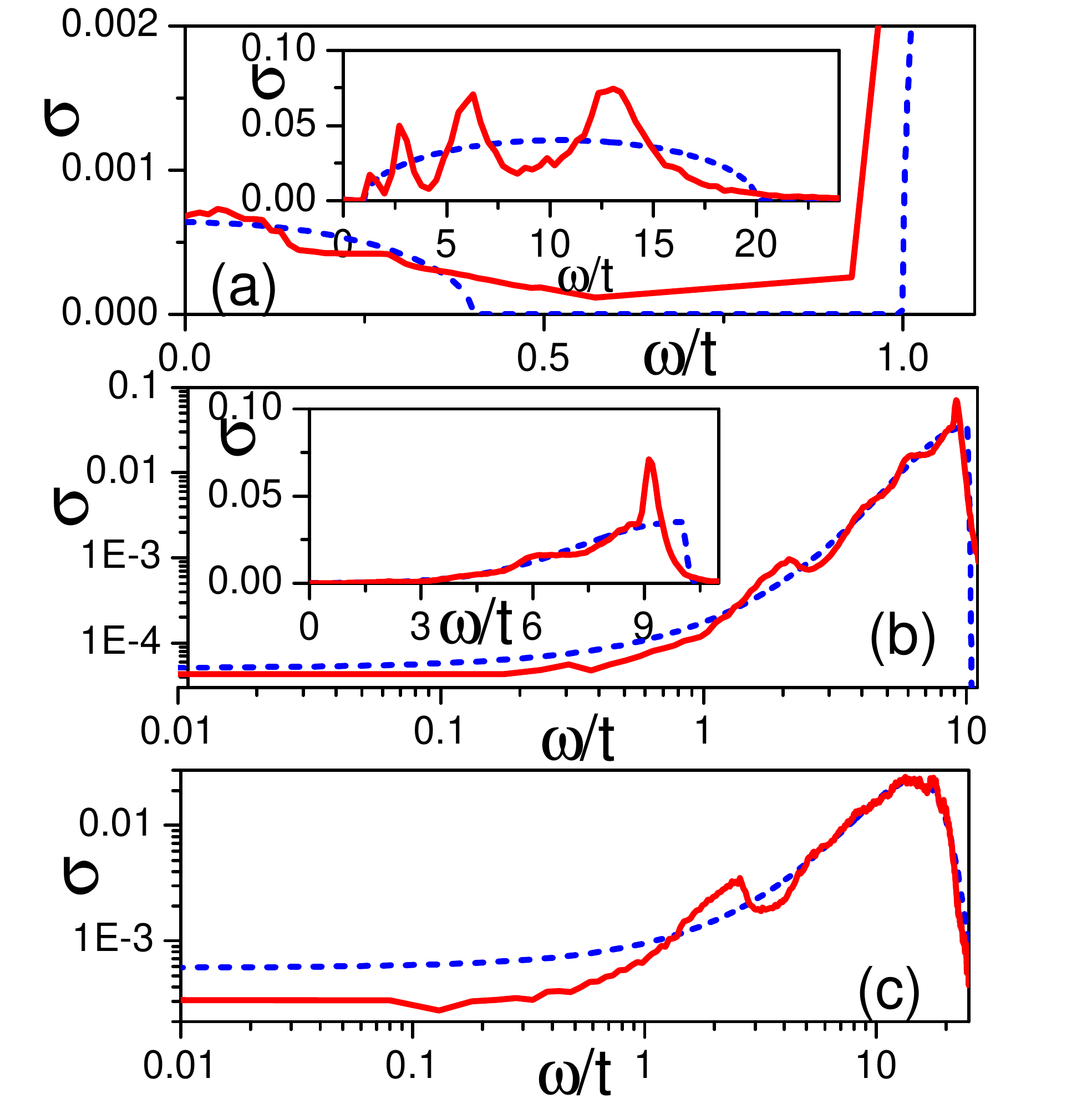}
\end{center}
\caption{\label{fig:fig4} Numeric tests for the reproducibility 
of the OC data at $\lambda=4$ and (a) $T=0.04$; (b) $T=0.32$; (c) T=0.5. 
Ideal spectra (dashed lines) are compared with spectra obtained through spectral 
analysis (solid lines).}  
\end{figure}

The last significant errorbar is the 
{\it systematic errorbar} $\delta \xi^{\mbox{\scriptsize {(sys)}}}$
which questions the ability of analytic continuation to give the correct value of mobility
in principle. 
To perform the test one takes the model function $\sigma(\omega)$, similar 
to the result $\widetilde{\sigma}(\omega)$ obtained for given parameters $\lambda$ and $T$, 
and considers it as a true spectrum.
Then, using Eq.~(\ref{s1}), one calculates $\widetilde{\Pi}(\tau)$, and tries to restore the value 
of mobility by solving again Eq.~(\ref{s1}). 
Indeed, this errobar is unique for every shape of the optical conductivity and 
temperature, is difficult to predict, and, thus, the procedure has to be repeated for all $\lambda$ 
and $T$ values.    

Figure~\ref{fig:fig4} shows several examples of comparison between true spectra and spectra
obtained by SOM procedure. 
Panel (a) shows spectrum which contains an high energy feature with large weight 
(see inset in panel (a)) and a low energy feature with tiny weight.
One can see that the value of mobility is reproduced with rather small errobar. 
To the contrary, the curves with high energy large weight peaks and featureless low 
energy continuum (Fig.~\ref{fig:fig4}(b-c)) point out the tendency to very large systematic 
errobars.    

Indeed, the flaw of the way to analyze the systematic errorbar lies in the fact that one 
does not know the {\it real} 
true spectrum and uses as true spectrum the OC {\it obtained} through the analytic continuation 
procedure. However, it is the best method that we are able to choose when the {\it real} spectrum is
unknown. Moreover, we found in numeric experiments that the relative systematic errorbar is 
reasonably  stable if we change the height of the model function $\sigma(\omega \to 0)$ by a factor 
of two. 
Its value mostly depends on the gross features of the spectrum, e.g. small structureless background
at low frequencies and large peak at high frequencies, etc. 
The last fact insures us that the value of systematic error is reliable even in absence of exact 
informations on the {\it real} spectrum because it is known that the most of the procedures of 
analytic continuations reproduce the gross features of the spectrum correctly.

\end{document}